\documentclass[sn-basic]{sn-jnl}
\usepackage{graphicx}
\usepackage{multirow}
\usepackage{amsmath,amssymb,amsfonts}
\usepackage{amsthm}
\usepackage{mathrsfs}
\usepackage[title]{appendix}
\usepackage{xcolor}
\usepackage{textcomp}
\usepackage{manyfoot}
\usepackage{booktabs}
\usepackage{algorithm}
\usepackage{algorithmicx}
\usepackage{algpseudocode}
\usepackage{listings}

\theoremstyle{thmstyleone}%
\theoremstyle{thmstyletwo}%

\theoremstyle{thmstylethree}%

\begin{document}

\title[Fast Radio Bursts]{The discovery and significance of fast radio bursts}

\author*[1,2]{\fnm{Duncan R.} \sur{Lorimer}}\email{duncan.lorimer@mail.wvu.edu}

\author[1,2]{\fnm{Maura A.} \sur{McLaughlin}}\email{maura.mclaughlin@mail.wvu.edu}
\equalcont{These authors contributed equally to this work.}

\author[3]{\fnm{Matthew} \sur{Bailes}}\email{mbailes@swin.edu.au}
\equalcont{These authors contributed equally to this work.}

\affil*[1]{\orgdiv{Department of Physics \& Astronomy}, \orgname{West Virginia University}, \orgaddress{\street{White Hall}, \city{Morgantown}, \postcode{26501-6315}, \state{WV}, \country{USA}}}

\affil[2]{\orgdiv{Center for Gravitational Waves and Cosmology}, \orgname{West Virginia University}, \orgaddress{\street{Chestnut Ridge Research Building}, \city{Morgantown}, \postcode{26506}, \state{WV}, \country{USA}}}

\affil[3]{\orgdiv{Centre for Astrophysics and Supercomputing}, \orgname{Swinburne University of Technology}, \orgaddress{\street{Hawthorn}, \postcode{3122}, \state{Victoria}, \country{Australia}}}

\abstract{In 2007 we were part of a team that discovered the so-called ``Lorimer Burst'', the first example of a new class of objects now known as fast radio bursts (FRBs). These enigmatic events are only a few ms in duration and occur at random locations on the sky at a rate of a few thousand per day. Several thousand FRBs are currently known. While it is now
well established that they have a cosmological origin, and about 10\% of all currently known sources have been seen to exhibit multiple bursts, the origins of these enigmatic
sources are currently poorly understood. In this article, we review the discovery of FRBs 
and present some of the highlights from the vast body of work by an international community. Following a brief overview of the scale of the visible Universe in \S 1, we describe the 
key moments in radio astronomy (\S 2) that led up to the discovery of the Lorimer burst (\S 3). Early efforts to find more FRBs are described in \S 4 which led to the discovery of the first repeating source (\S 5). In \S 6, as we close out on the second decade of FRBs, we outline some of the many open questions in the field and look ahead to the coming years where many surprises are surely in store. }

\keywords{fast radio bursts, compact objects, radio astronomy, transient astrophysics}



\maketitle

\section{Introduction to the Universe}

For millennia humans have gazed upon the night sky and pondered what our place is in the Universe and what laws govern it? The unaided human eye can only see about 6000 stars but for many years their distances, lives and death were a mystery. It wasn’t until astronomers applied mathematics to observational data that our understanding of the Universe and the night sky started to { progress rapidly}. Following the Copernican Revolution which inspired Tycho Brahe's precision measurements of the positions of the planets, Johannes Kepler used Brahe’s work  to understand their trajectories, and developed his famous three laws of planetary motion. A further revolution in our understanding of the Universe occurred when Galileo made the first recorded observations of the night sky with a telescope in the early 1600s. His observations revealed that Jupiter was orbited by four major moons, and this displaced the prevailing view that Earth was the centre of the Universe. Four centuries of observations have followed, and the astronomical and theoretical breakthroughs are astonishing. The heavens have changed from a static background with pinpricks of light and wandering planets to a dynamic and violent Universe where phenomenal instruments are revealing the innermost workings of complex systems involving stars, planets, black holes, white dwarfs and neutron stars.

Our Earth is only 40,000~km in circumference, but a million of them would comfortably fit inside our Sun. Fortunately the Sun has almost constant warmth, with an estimated remaining lifetime of about 5 billion years, and this has enabled the evolution of life on Earth. But the Sun is a fairly average star in the Universe, and will eventually puff off its outer layers when it becomes a red giant leaving behind a white dwarf, the ashes of its core. Some stars are much more massive than the Sun and have more violent deaths. Rigel, for instance, is about 20 times the mass of the Sun and will die in a violent supernova explosion, leaving behind a neutron star. Neutron stars are only about 10~km in radius but weigh over half a million Earth masses. A teaspoon of a neutron star material weighs more than the whole of humanity. Even larger stars collapse to form black holes, and these can be in excess of 50 solar masses. Black holes are so dense that not even light can escape their extreme gravity.

In 1054~AD Chinese astronomers witnessed the supernova explosion that created the Crab Nebula when a massive star like Rigel died. For a few weeks the remnant was so bright it could be seen during the day. At the heart of the nebula was a neutron star, the famous Crab pulsar. It rotates every 33~ms and emits radiation all across the electromagnetic spectrum, from radio waves to gamma-rays.

The Milky Way is full of the remnants of exploded stars, and comprises $\sim 400$ billion suns, as well as gas and dust. But the Milky Way is only one galaxy in the Universe, which is thought to contain many 10s--100s of billions of galaxies of different sizes. Our Milky Way has a pancake-like structure with a central bulge and bar, and spiral arms. It is very flat in one dimension, and extends for something like 100,000 light years in diameter. Its halo is sparsely populated by stars that have been ejected from the disk and is mainly composed of older stars. It is referred to as a spiral galaxy.

One of our closest neighbours is the Andromeda galaxy, similar in shape and size to the Milky Way. Other galaxies lacked the spin to collapse into spiral galaxies, and have a more ellipsoidal shape. Their stars tend to be ``red and dead'', as their star formation era has long-since passed. Astronomers call these galaxies ``ellipticals''. Although less luminous, smaller fragments of galaxy formation abound, and our two nearest neighbours are known as the Magellanic Clouds. These small, irregular galaxies often exhibit a haphazard shape, are still producing new stars and orbit the Milky Way every billion years or so. They are archetypal ``irregular'' galaxies.

Astronomers have mapped the distribution of galaxies in the wider universe and find that they are consistent with numerical simulations of a once very smooth universe that has both expanded and had gas clouds collapse to form galaxies. The galaxies cluster along filaments and form groups. Sometimes they collide, triggering new waves of star formation. At the cores of galaxies, supermassive black holes accrete matter and spew forth relativistic particles in powerful jets. These are known as quasars \citep{1963Natur.197.1040S}.

Between the galaxies the density of atoms drops to only about one per cubic metre, over 10,000 times less dense than that of galaxies. Light and radio waves traverse the enormous distances across the Universe with a small probability of interaction. This property has enabled astronomers to see the Universe age by looking back in time. On its long journey, visible light largely ignores the ionised gas between the galaxies, making its composition difficult to study.
Modern astrophysics is driven by our desire to learn about what classes of objects exist, such as planets, white dwarfs, stars, quasars, neutron stars and black holes, but also how they live and die, and how they can be detected. Astronomers also attempt to measure the mass and composition of the Universe, its age and dimension, as well as what physical laws are at play. It turns out that fast radio bursts have an important role to play in developing our broader understanding of the cosmos.

\section{The transient radio sky}

Although optical astronomy has existed since humans first gazed at the heavens, radio astronomy is less than 100 years old. The first radio telescopes were not sensitive to short timescale phenomena; they focused on making maps of the radio sky that were unchanging and static. With the discovery of quasars in 1964, however, that changed. Because the angular sizes of quasars are so small, quasars are observed to ``scintillate'', or twinkle as their radio waves travel through the ionized interplanetary medium.

\begin{figure}[h]
\centering
\includegraphics[width=0.9\textwidth]{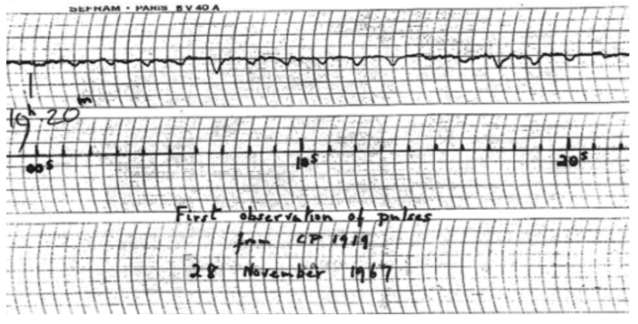}
\caption{Pen chart recording showing the detection of pulses from the original pulsar B1919+21 first seen during the Cambridge survey~\citep{1968Natur.217..709H}.}\label{fig1}
\end{figure}

In 1967, Jocelyn Bell Burnell was a graduate student working with Prof. Anthony Hewish at Cambridge University in the UK. Her thesis project was to study this phenomenon of quasar scintillation \citep{1969PhDT.......162B}. To do this, she used a chart recorder that recorded the intensity of radio emission over the sky to search for the types of variations expected from scintillating quasars. One day, Bell Burnell saw a signal that occurred on a much shorter timescale than expected for quasar scintillation. It also did not look like human-made radio frequency interference, and it appeared at the same local sidereal time each day, meaning that it was coming from space.  Soon, Bell Burnell and Hewish found that the signal consisted of regularly spaced pulses, with a separation of exactly 1.337 seconds, as shown in Figure~1. This short timescale meant that the signal must be coming from a very small object. Bell Burnell and Hewish jokingly nicknamed it “LGM 1”, or “Little Green Man 1”, as this period was too short to come from anything like a normal star and could have been from aliens~\citep{1968Natur.217..709H}.

Soon Bell Burnell and Hewish found more of these objects, in different parts of the sky, confirming that they were a new class of astronomical sources. Following the suggestion of a journalist, these ``pulsating radio sources'' were henceforth referred to as pulsars. Not long after, Thomas Gold and Franco Pacini both suggested that these pulses were neutron stars born in the supernova explosions of massive stars. Soon thereafter, when the 33-ms Crab pulsar was discovered at the centre of the Crab supernova remnant in 1968~\citep{1968Sci...162.1481S} and the 89-ms pulsar was found at the centre of the Vela supernova remnant~\citep{1968Natur.220..340L}, this theory was confirmed. 

Pulsars are fascinating objects with extreme properties. Because of the conservation of angular momentum, they rotate very rapidly, with spin periods ranging from 1.4 milliseconds to tens of seconds. They also have extremely high magnetic field strengths of $10^8$ to $10^{14}$ Gauss, trillions of times higher than the magnetic field of the Earth. They are extremely dense, with masses up to twice as high as the Sun but with radii of only 10~km. This is an equivalent density to taking all the people on Earth and squeezing them into a thimble! Pulsars’ rapid spin periods and high magnetic fields lead to acceleration of particles above their magnetic poles, and this produces beamed radio emission. Pulsars therefore act like interstellar lighthouses; every time the pulsar beam swings past our line of sight, we detect a pulse.

One very important property of pulsar emission is that it is dispersed. This means that the lower frequencies of the radio pulse arrive later than the higher frequencies. The time delay, $\Delta t$, between a pulse at a high frequency, $\nu_{\rm hi}$, compared to a lower frequency, $\nu_{\rm lo}$, is given approximately by
\[
    \Delta t \simeq 4150~{\rm s} \times \left[
    \left(\frac{\nu_{\rm lo}}{\rm MHz}\right)^{-2} -
    \left(\frac{\nu_{\rm hi}}{\rm MHz}\right)^{-2}
    \right] \times \left(
    \frac{{\rm DM}}{{\rm cm}^{-3}~{\rm pc}}
    \right),
\]    
where the ``dispersion measure'' DM is the integrated column density
of free electrons along the line of sight to the source.
Dispersion occurs due to interactions of the radio photons with electrons and other charged particles in the interstellar medium, the gas in between the pulsars and Earth. On one hand, dispersion is a nuisance, as it means we must correct for this sweep before searching for pulsars in radio data. On the other hand, it is a valuable tool, as the amount of this dispersive sweep, coupled with a model for Galactic electron density, gives us a handle on how far away a pulsar is. We have estimated distances to many pulsars through this method.

After the discovery of the first pulsars, astronomers soon realised that searches that were sensitive to a pulsar’s time-averaged emission were much more efficient and sensitive that those that searched for individual, single pulses. In the 1970s, searches that used Fourier transforms to measure excess power at particular rotation frequencies became commonplace~\citep{1969A&A.....2..280B}, and astronomers generally stopped searching data for the single, dispersed pulses through which the first few pulsars were discovered.
Despite this, one special class of pulsars could be detected with higher sensitivity through their individual single pulses - giant pulsing pulsars. These pulsars, of which the Crab pulsar is the first-discovered and most well-known example, emit single pulses that are occasionally 100s or even 1000s of times brighter than the average pulse. These pulses are very narrow and occur randomly in time, as shown in Figure~2.

\begin{figure}[h]
\centering
\includegraphics[width=0.9\textwidth]{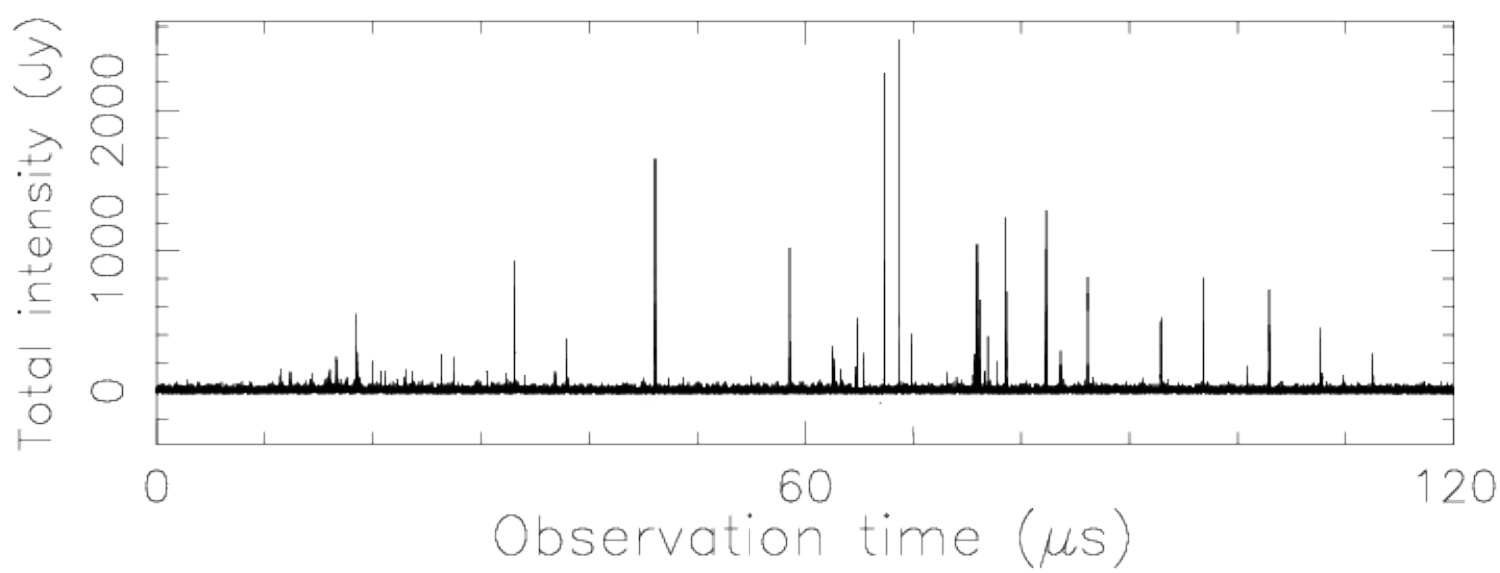}
\caption{This observation, { carried out in the 5--6~GHz radio frequency band}, is among the highest time resolution views of emission seen in the radio sky so far~\citep{2003Natur.422..141H}. These nanosecond scale pulses are being emitted by the Crab pulsar. Credit: Tim Hankins.}\label{fig2}
\end{figure}

The discovery of the Crab pulsar using the 300~ft Green Bank telescope through its giant pulses~\citep{1968Sci...162.1481S} and subsequent identification of the 33~ms period using the 1000~ft Arecibo telescope \citep{1969Natur.221..453C} generated significant interest in the prospect of finding extragalactic sources similar to the Crab. Following prophetic
remarks by \citet{1970QJRAS..11..443B}, \citet{1971ApJ...165..509C} and \citet{1972Ap&SS..19..431C} discussed some ideas about emission mechanisms for such sources. Also on
the theoretical side, \citet{1973Natur.246..415G} highlighted the prospects for using extragalactic sources to map out the electron content of intergalactic space in a similar way to what was then being done with the growing sample of pulsars to map out the interstellar medium of the Milky Way~\citep{1969ApL.....4..229M}.

Following a number of attempts to find extragalactic radio pulses at radio wavelengths \citep{1973BAAS....5Q.284C,1974ASSL...45...61J,1974ApJ...187L..57H,1975ApJ...201L.113C,1979Natur.277..117P,1981Ap&SS..75..153C}, \citet{1980ApJ...236L.109L} announced the discovery
of radio pulses detected from the direction of the giant elliptical galaxy M87. However,
in spite of a number of follow-up searches \citep{1981ApJ...244L..65T,1981ApJ...244L..61H,1981ApJ...245L..99M,2021ApJ...920...16S}, no further radio pulses were ever seen from M87. While it was never conclusively determined
whether the events seen by Linscott and Erkes were from highly variable sources in M87 or
some form of interference, we note that the galaxy is strongly suspected to 
contain detectable sources \citep{2023ApJ...944....6K}. Large-area searches for extragalactic radio pulses in the intervening years continued to yield
null results \citep{1989PASA....8..172A,1998A&A...329...61B,2003PASP..115..675K}. Meanwhile, \citet{1993ApJ...417L..25P} had emphasized the importance of
dispersion in the detection of cosmological radio transients and searches
for `prompt' radio emission from gamma-ray bursts were being carried out
\citep[see, for example,][]{1995Ap&SS.231..281G,1999PhDT.........3B}

As a graduate student at Cornell, one of the authors (Maura McLaughlin) and her advisor Jim Cordes had a goal to detect pulsars in other galaxies. At that time, nearly all of the pulsars known were in the Milky Way, with just a handful in the Small and Large Magellanic Clouds, nearby satellite galaxies to the Milky Way. \citet{2003ApJ...596..982M} realised that probably the only way to detect pulsars in galaxies further than these was to search for extremely bright giant pulses, as opposed to using the Fourier transform searches that were almost exclusively used for searches at the time. They developed methodology and code that they applied to a search to the local group galaxy M33 (the Triangulum galaxy), which is roughly 1~Mpc away. While they { found no bursts}, they had developed the tools that could be used to carry out other searches for single pulses.

A breakthrough in the field came shortly after the Arecibo searches when, 
applying this single-pulse code to data from the Parkes Multibeam Pulsar Survey, 
\citet{2006Natur.439..817M} found roughly a dozen pulsars that were not detectable in previous Fourier transform searches. These objects, which the authors nicknamed ``rotating radio transients'' (RRATs), emitted a very small number of detectable pulses over the course of the 35-min search observations, meaning they could not be detected through a Fourier transform search but could be through a search for single pulses. After the discovery of the RRATs, pulsar astronomers world-wide began to use the single-pulse search technique routinely as part of pulsar searches. Despite their sporadicity, the properties of the RRAT single pulses were not significantly different from the properties of normal pulsar single pulses. However, looking at the properties of these objects in this way raised a tantalising question: what other types of objects could exist in this newly revealed single-pulse phase space?

\section{Discovery of the Lorimer Burst}

While all the work that led up to this point prepared us to find the first fast radio burst, it took the act of applying these tools analysing archival datasets to make the discovery. Shortly after two of the authors (Maura McLaughlin and Duncan Lorimer) arrived at West Virginia University (WVU), as assistant professors of physics in 2006, they began to work with students on a variety of projects. One of these was to look through recently published data from Parkes looking at the Magellanic Clouds \citep{2006ApJ...649..235M}.
The Parkes data had already been extensively searched for periodic pulses from pulsars, but represented a self-contained experiment as a place to look for bright individual pulses. At that time, computational resources were sufficiently developed for it to be feasible to process the data on a desktop computer over a period of a few months. This was adequate to allow then undergraduate student Ash Narkevic to manually inspect the results of the data analysis at a pace that worked well with their studies within an academic year. The diagnostic plots were essentially representations of the data as a function of dispersion measure, DM, versus time. It was Ash’s task to look for signals that were statistically significant and occurred at non-zero DM. Fortunately the Parkes data set contained a number of known pulsars in the Magellanic Clouds that were bright enough such that they could be seen as individual pulses as shown in Figure~3.
\begin{figure}[h]
\centering
\includegraphics[width=0.9\textwidth]{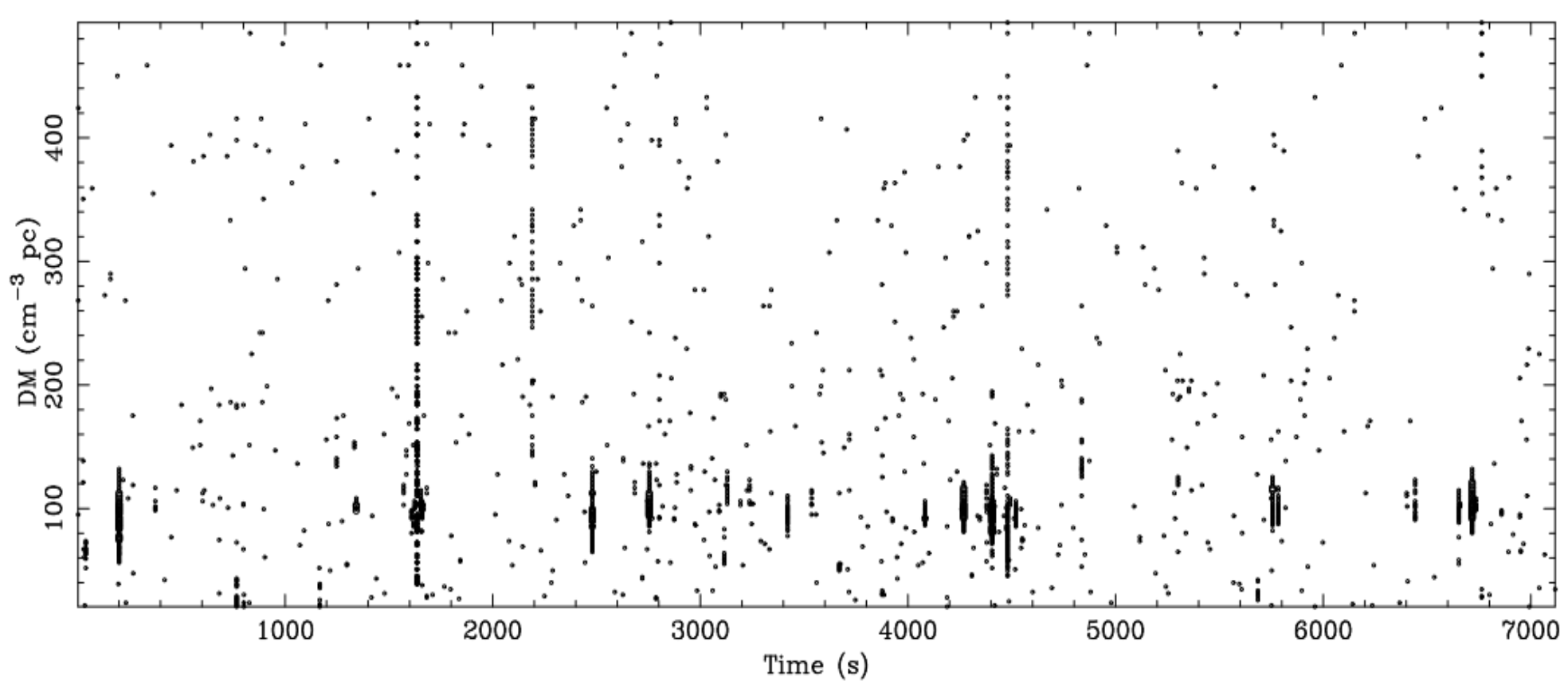}
\caption{Single-pulse search output showing individual pulses from PSR~B0529--66 in the Large Magellanic cloud { observed in the 1.2--1.5~GHz frequency band}. In this visualisation, the pulses are seen as the discrete vertical stripes at a constant DM of 100~cm$^{-3}$~pc.}\label{fig3}
\end{figure}
Ash would meet with Duncan for an hour once a week to review findings from the previous week’s processing. Most of the signals were relatively straightforward to categorise: a form of interference, an already known pulsar in the data, or something that perhaps { was} right at the edge of the sensitivity threshold, and too faint to confidently assert the presence of a signal. One day, Ash found something that didn’t fit these categories. The discovery plot is shown in Figure~4.

\begin{figure}[h]
\centering
\includegraphics[width=0.9\textwidth]{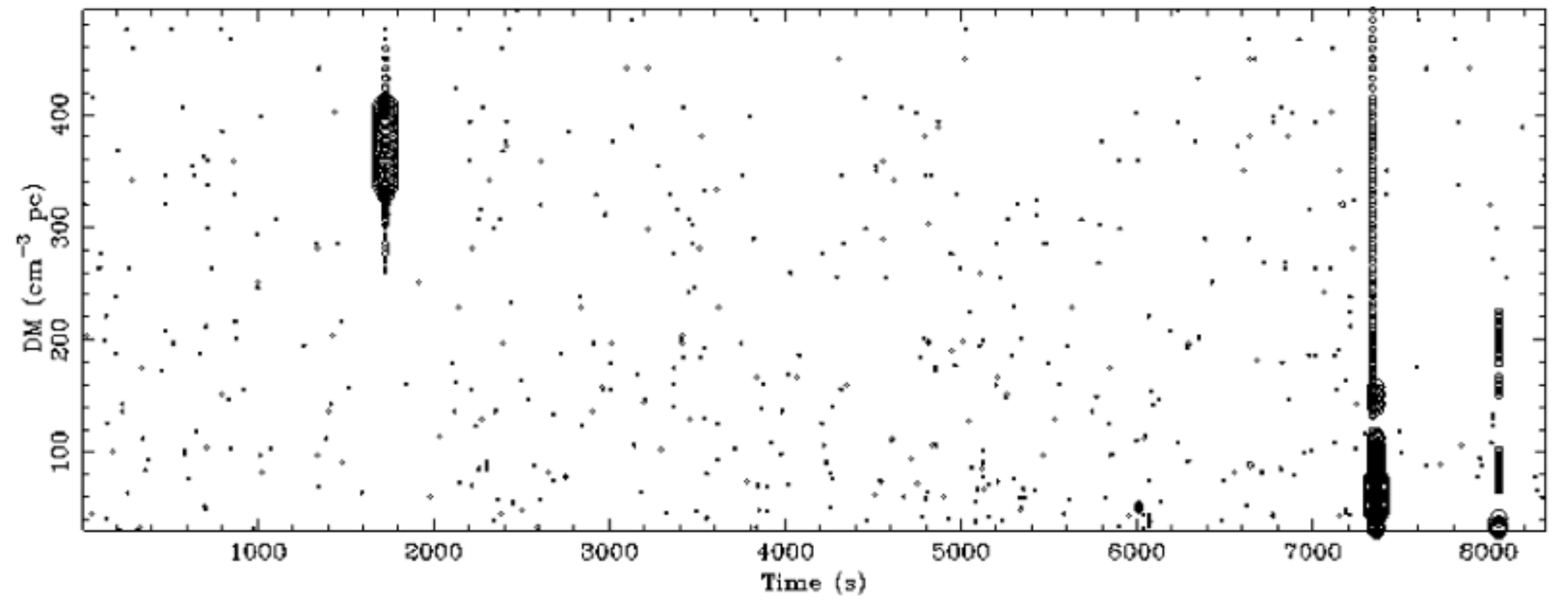}
\caption{Single-pulse search output showing the Lorimer burst which appears as the bright localised signal at a DM of 375~cm$^{-3}$~pc. { These data were collected in the 1.2--1.5~GHz frequency band.}}\label{fig4}
\end{figure}

As can be seen, the presence of a bright individual pulse is seen at a DM of 375~cm$^{-3}$~pc at around 1750~s after the start of the observation. The only other signals present in this graph are the vertical stripes seen at 7400 and 8000~s; the vertical stripes seen for these are the hallmarks of strong locally generated interference. The new signal was very well defined in DM space, as shown by the increasing and decreasing widths of the symbols in Figure~4. Further inspection of this signal showed that the sky position was significantly offset from the Small Magellanic Cloud. This observation had been collected as part of an effort to cover both the clouds using the Parkes multibeam receiver. This 13-pixel radio camera allows users to observe larger fields than possible using a single beam that was available to most telescopes back at that time. Because of the way the field of the Small Magellanic Cloud was covered, this particular pointing was looking mostly away from the cloud itself and into the distant Universe.  The DM that we observed for this pulse about 10 times the expected value from the Milky Way in that direction. The pulse appeared to be firmly extragalactic in origin.

These initial checks also showed the presence of the signal in one other pixel in the multibeam system. During a visit to Parkes in Spring 2007, Duncan shared this discovery with the third author of this paper (Matthew Bailes). Upon even further scrutiny while they were together, it was realised that the signal was strongly present in one further beam of the system. It was originally not registered by the code as the pulse was so bright that it saturated the digitizer levels in the system and the data preparation software had erroneously flagged that particular signal as interference! All in all, the pulse's signal can be  detected in 4 out of 13 of the multibeam pixels (one quite weakly) and it had an anomalously high DM. Visualisations of the pulse in terms of frequency versus time, along with an example pulse from a known pulsar with a similar DM are shown in Figure~5.

\begin{figure}[h]
\centering
\includegraphics[width=0.9\textwidth]{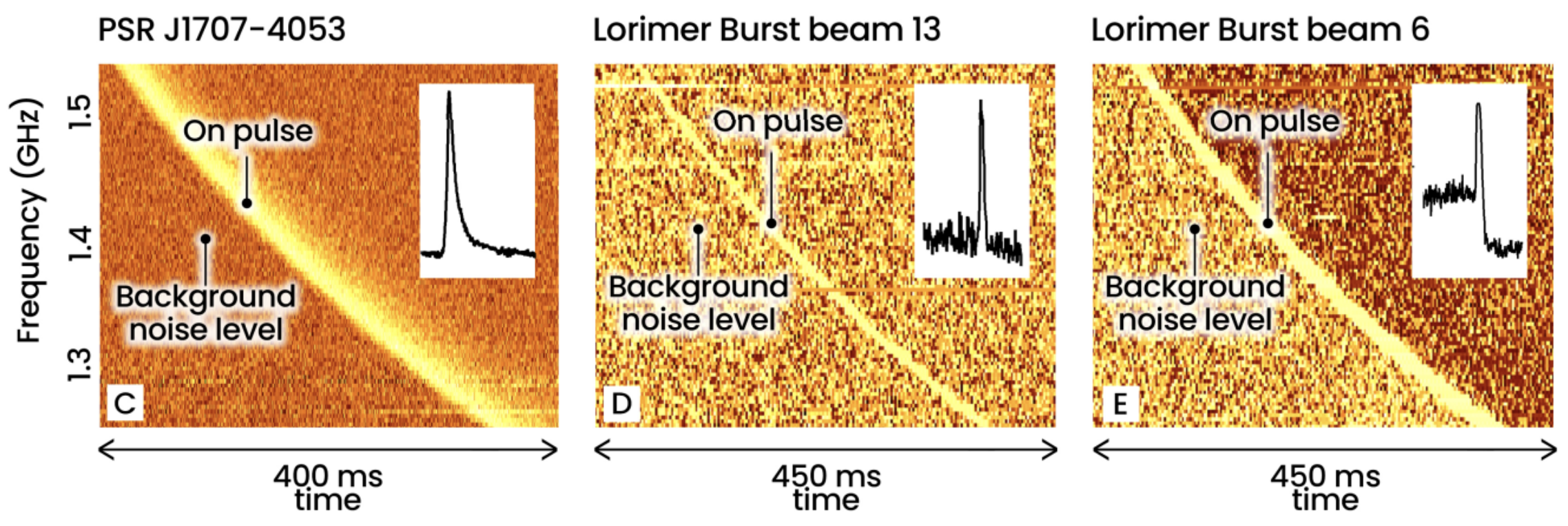}
\caption{Waterfall plots showing radio frequency versus time for the bright pulsar PSR~J1707--4053 (left) and the Lorimer burst seen in two beams of the multibeam receiver (centre and right).}\label{fig5}
\end{figure}

The dispersion of the pulse provides a strong constraint on its distance. After subtracting off the expected DM from the Milky Way, there is about 330~cm$^{-3}$~pc of “extragalactic dispersion” produced mostly by the free electrons in the intergalactic medium. Based on results from other astronomical surveys, it is known that the density of the IGM is roughly one electron per cubic metre. A back-of-the-envelope calculation (DM divided by mean density) then gives an implied distance of 330~Mpc. Using slightly more sophisticated calculations~\citep{2007Sci...318..777L}, we suggested that the distance could be as high as 1~Gpc. It appeared that the pulse originated from a source at cosmological distances. In addition to the observed dispersion, we also were able to see an evolution of the pulse width with frequency, with the lower-frequency components of the pulse being more scattered than their higher-frequency counterparts. The morphology of the pulse was consistent with its propagation through a turbulent ionised medium.

\begin{figure}[h]
\centering
\includegraphics[width=0.8\textwidth]{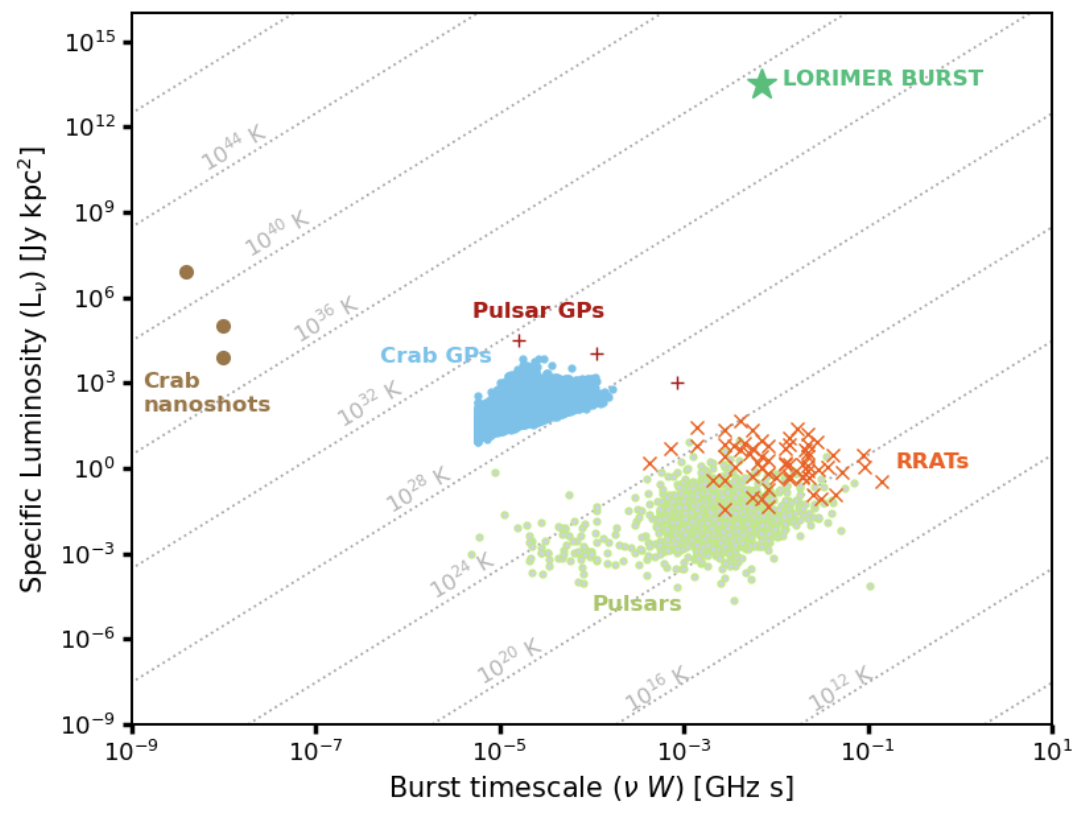}
\caption{Individual pulses from a variety of different coherent radio sources displayed as luminosity versus pulse width times observing frequency. The highly luminous nature of the Lorimer burst made it clearly anomalous at the time of its discovery. Figure credit: Chris Flynn and Manisha Caleb.}\label{fig6}
\end{figure}

Given such a high implied distance, the inferred luminosity of the new source, colloquially known as the ``Lorimer burst'', was phenomenally bright. As shown in Figure~6, it was something like 10 to 12 orders of magnitude brighter than anything known at the time. The Lorimer burst was apparently the prototype of a new class of distant radio transient source. Very much like the discovery of pulsars some 40 years earlier, this serendipitous discovery had a wealth of implications some of which were outlined in the discovery paper \citep{2007Sci...318..777L}.

Beyond its anomalous luminosity, we realised that (like pulsars) the short pulse duration implied a coherent emission mechanism with a very high brightness temperature. While considerably uncertain due to the lack of a direct measurement of distance to this source, we inferred source energetics of the order of $10^{33}$~J. Light travel time arguments limited the size of the emitting region to be less than 1500~km. Not only was the source very small, bright and very far away, our observations to date implied an all-sky rate of hundreds of similar sources per day over the sky. Finally, in spite of over 90~hr of follow-up, no further bursts were seen from this location. 

Taken together, these lines of evidence suggested a source mechanism that was catastrophic in nature --- perhaps a neutron star inspiral, supernova, or gamma-ray burst. With statistics of one object, we realised that these ideas were speculative at best. Nevertheless, in our paper and accompanying cover letter to the editor of Science, we boldly posited this source as a prototype and eagerly encouraged further searches of other archival datasets by ourselves and other groups.

The months following the discovery publication in Science were incredibly exciting. There was a lot of interest in both the popular press and the astrophysical community. Naturally, many questions were asked. Why was the source so bright? Where was the more numerous population of fainter sources? In spite of our best arguments in favour of an astrophysical origin, could the source be due to something else that we had not contemplated? A nearby anomalously dispersed object, or some local interference? Only further follow-up observations could really answer these and other questions.

\section{Early follow-ups}

Radio astronomers soon pored through old radio surveys of the sky to look for further examples of the Lorimer Burst, and also commenced new surveys looking further into the universe for them. We had observed the location of the Lorimer Burst for 90 hours, but not seen any repeat bursts. New surveys failed to find other examples, and there was some alarm when one of Matthew’s PhD students (Sarah Burke-Spolaor, now a faculty member at WVU) found some sources of interference \citep{2011ApJ...727...18B} at the Parkes site with dispersion characteristics bearing some similarities to the Lorimer Burst. These were dubbed “Perytons” named after the mythical Greek elk that casts a human shadow. These mysterious sources cast doubts on the celestial nature of the Lorimer Burst. The Perytons were imperfectly dispersed signals, clearly close to the telescope site, and worryingly close to the Lorimer Burst’s DM of 375~cm$^{-3}$~pc.

For over five years no further extragalactic radio bursts were identified, and the community started to express scepticism as to their reality. Hopes resurfaced when a team led by Evan Keane \citep{2012MNRAS.425L..71K} found a putative burst near the plane of our Galaxy, but it was unclear whether it was part of our own galaxy’s stellar population or not due to our imperfect knowledge of the density of the interstellar medium along the Galactic plane.

Just prior to the discovery of the Lorimer Burst, the High Time Resolution Universe (HTRU) collaboration had formed with the aim of using new digital instrumentation to more effectively search for millisecond pulsars in the Milky Way with the Parkes radio telescope~\citep{2010MNRAS.409..619K}. The new digital instrumentation had about 8 times the frequency resolution of the old Parkes filterbanks, higher time resolution and up to 8-bit resolution.

Dan Thornton was a PhD student under the supervision of Ben Stappers at the University of Manchester \citep{2013PhDT.......212T} and was assigned the search for Lorimer Bursts by the collaboration. His initial efforts were proving frustrating, as radio interference was masking any putative bursts. His advisor suggested he just set a high detection threshold, and, remarkably, a 50$\sigma$ burst was detected near a DM of 1000~cm$^{-3}$~pc well away from the Galactic plane. The new digital instrumentation allowed very precise measurement of the burst's dispersion and scattering time scale power law indices, which were very near the expected values of --2 (dispersion) and --4 (scattering) expected for celestial sources. Spurred on by his discovery, Thornton soon found another three examples, all with different dispersion measures, all far from the Galactic plane, and all with vertical components of dispersion measures well in excess of any known radio pulsar. The cosmological population of radio bursts had been discovered \citep{2013Sci...341...53T}. The HTRU collaboration suggested the name “fast radio bursts” or FRBs, a name that has stuck to this day.
These discoveries spurred many different teams into action. Laura Spitler and collaborators  soon discovered \citep{2014ApJ...790..101S} the first FRB with the Arecibo telescope in archival data. It had occurred back in 2012 and was dubbed FRB 121102 (on UTC date 2 Nov 2012). Matthew was greatly relieved to see a burst at another facility other than Parkes, and when she previewed its waterfall (dispersion) plot at Manchester he felt like leaping on stage and hugging her for joy!

Parkes remained a fruitful source of FRBs, the HTRU collaboration published another 5, one with a DM beyond 1600~cm$^{-3}$~pc that appeared to consist of two components, and other groups started to find them both at Parkes and other facilities like the Green Bank Telescope~\citep{2015Natur.528..523M}. The GBT burst possessed a high rotation measure, consistent with it emerging from a magnetised environment.

At the Molonglo Observatory, \citet{2017PASA...34...45B} developed machine learning algorithms to identify bursts in real time and enable post facto coherent dedispersion that revealed that FRBs could consist of multiple narrow components, with timescales less than a few 10s of microseconds \citep{2018MNRAS.478.1209F}.
These small timescales meant that the source size could only extend tens of kilometres and pointed to an origin in the magnetospheres of neutron stars. One Parkes FRB had a dispersion measure of over 2000~cm$^{-3}$~pc and was potentially from a source so distant that the Universe was much less than half of its current age when it was emitted. The field was however transformed when routine follow-up observations of the only known Arecibo burst were conducted in 2015.

\section{The repeating FRB --- a ``minor point of interest''}

All of the FRBs discovered up until the time of the Arecibo FRB had been subject to many hours of follow-up observations with Parkes and other telescopes, with no repeated bursts found. It was becoming almost generally accepted that these were “one-off” phenomena when additional bursts were discovered from FRB~121102. A { Canadian graduate} student working { under the supervision of Vicky Kaspi at McGill University} on the Arecibo survey named Paul Scholz { jokingly} titled the email announcing this result in November 2015 as “A minor point of interest regarding the Spitler burst”. These bursts occurred at the same DM and sky position as the original FRB, and so there was no doubt they were associated. As more and more bursts were detected from this FRB, it was clear that they showed an astonishing diversity of properties. Some were narrow, some broader. Some peaked in brightness at high frequencies, and some at low frequencies. Some had a single pulse component, while some appeared to have multiple components \citep{2016Natur.531..202S}. { Further multi-wavelength observations of this source soon followed \citep{2016ApJ...833..177S} and it was shown that the arrival time distribution of the radio 
bursts was highly non-Poissonian.}

\begin{figure}[h]
\centering
\includegraphics[width=0.8\textwidth]{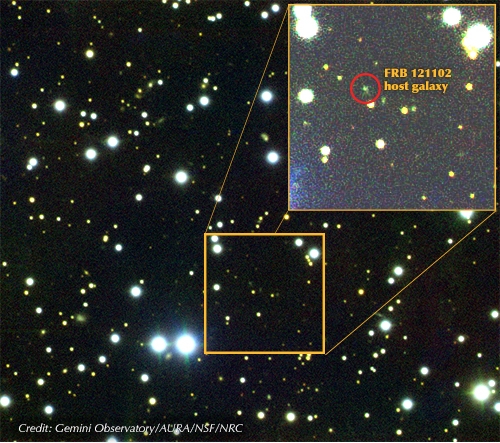}
\caption{Follow-up of FRB~121102 showing the identification of a host galaxy at $z=0.19$ (inset).}\label{fig7}
\end{figure}

The repetition of this FRB made it possible to observe its position with an interferometer to localise it much more precisely than possible with a single-dish telescope like Arecibo. Observations with the VLA by \citet{2017Natur.541...58C} revealed that the burst was associated with a galaxy. This was very important for several reasons. First, these bursts were definitely astrophysical and coming from stellar populations. Second, as shown in Fig.~\ref{fig7}, the galaxy was detected optically { by \citet{2017ApJ...834L...7T}}, providing a measurement of the redshift, $z=0.19$. Hence, for the first time the DM of a burst could be used to constrain the electron density of the intergalactic medium. Third, the kind of galaxy it came from provided important clues to its origin. In this case, FRB~121102 originated in a dwarf galaxy with a high star formation rate, indicating that the burst came from stellar populations such as neutron stars. It was also found that the bursts from FRB~121102 had one of the highest rotation measures ever observed~\citep{2018Natur.553..182M} which means it came from a very magnetic, dense environment.

After the discovery of this FRB, each year brought more and more FRB discoveries, with this number exploding once the CHIME telescope in Canada came online in 2018~\citep{2018ApJ...863...48C}. This telescope began discovering FRBs at a very high rate, and there are currently 536 CHIME-discovered FRBs in the public domain~\citep{2021ApJS..257...59C} with several thousand more to be released soon. As more FRBs began to be detected, more puzzles emerged. { Although around 10\% of all currently known FRBs exhibit repeat bursts, an analysis of the CHIME sample shows that the true repeater fraction is probably closer to aroud 3\%~\citep{2023ApJ...947...83C}.}
One of the repeating FRBs detected with CHIME appeared to show a 16-day periodicity in its active cycle~\citep{2020Natur.582..351C}. More recently, there are hints that the original repeater FRB~121102 demonstrates a roughly 180 day periodicity~\citep{2020MNRAS.495.3551R}. Periods of this timescale could be indicative of binary motion, and several models have been proposed. One attractive model involves modulation as a neutron star travels through the wind of its stellar companion~\citep{2022Sci...378.3043B}.

\section{Recent breakthroughs and the future}

The current state of the field of FRBs is very active. Barely a day goes by without a new paper appearing discussing either a new result, or trying to make sense of some of the many observational breakthroughs that are being presented. Among the recent breakthroughs is the discovery of an FRB-like pulse from a known magnetar in our Galaxy, SGR~1935+2154. This incredibly bright pulse was spotted by CHIME \citep{2020Natur.587...54C} { and subsequently also by the STARE2 experiment \citep{2020Natur.587...59B}} and quickly localised to a magnetar which was known to be undergoing an enhanced period of activity at the time (April 28, 2020). With a peak flux density of over 1~MJy, the equivalent luminosity of this pulse was several thousand times brighter than anything observed from the Crab pulsar, and only a factor of 30 fainter than a typical cosmological FRB pulse. It therefore appears to make a strong case for magnetars as plausible sources of FRBs. Magnetars have now long been known as sources of energetic emission and the 2021 Shaw Astronomy Prize\footnote{https://www.shawprize.org/laureates/2021-astronomy} Laureates, Prof.~Vicky Kaspi and Prof.~Chryssa Kouveliotou, were recognized for their contributions to our understanding of the magnetar population in the Galaxy.

A simple picture in which all FRBs originate from young magnetars, however, does not seem to be sufficient to explain the diversity of environments in which FRBs are being found. The sheer number of FRB discoveries being made by CHIME (around 3 per day on average) is leading to a variety of FRB locations. While we see one source located in a spiral arm of a nearby galaxy, a very intriguing discovery is FRB 20200120E in the globular cluster associated with the nearby galaxy M81 \citep{2021ApJ...910L..18B,2022Natur.602..585K}. Globular clusters are among the oldest stellar populations, and regions where young magnetars are not expected due to the lack of any recent star formation activity. As recently pointed out by \citet{2023MNRAS.525L..22K}, magnetars { and other young neutron stars} could form in these environments through the mergers of white dwarf binaries. { As discussed by \citet{2023MNRAS.525L..22K}, this could} also explain why our own Galaxy’s globular cluster system harbours a few young neutron stars which are otherwise very difficult to reconcile with the older population of millisecond pulsars currently seen. { Further observational and theoretical studies in this area are eagerly anticipated in the future.}

At the time of writing, March 2024, almost 50 FRBs have been conclusively associated with host galaxies. To date, there appears to be no statistically significant difference between the hosts of repeating versus non-repeating FRBs. In general, FRBs are found in galaxies where there are at least moderate amounts of star formation. They tend not to be found in the so-called “red and dead” galaxies where star formation has ceased~\citep{2022AJ....163...69B}. When tracked as a function of redshift, FRBs generally seem to trace the evolution of star formation in galaxies, at least out to redshifts around 0.5. A statistically significant sample at higher redshifts is currently lacking, with the highest redshift FRB to date being found at $z=1$ \citep{2023Sci...382..294R}.

One of the amazing aspects of FRB studies is that they act as probes of the intervening matter, even if one does not fully understand their source populations, which we currently do not!  All that is needed is a sample of FRBs with well-determined redshifts. From this, it is possible to effectively count the number of electrons along different sight lines in the Universe and measure the electron density directly. As one journalist, Joe Palca, points out: very much in the same way as we use dipsticks to keep track of the oil content of our cars, astronomers can in effect use FRBs as “intergalactic dipsticks” to measure the electrons in the universe. Although conceptually simple, this turns out to be a very challenging exercise. The first team to do it was led by J-P Macquart using a sample of FRBs that was very carefully obtained using the ASKAP telescope~\citep{2020Natur.581..391M}. This required the coordination of a large team of astronomers to get both the radio measurements and optical redshifts necessary. The result, shown in Figure~8, is now known as the {\it Macquart relation}, in honour of J-P’s memory. J-P passed away far too soon, but left the astronomical community with an amazing result first anticipated by \citet{1973Natur.246..415G}.

\begin{figure}[h]
\centering
\includegraphics[width=0.8\textwidth]{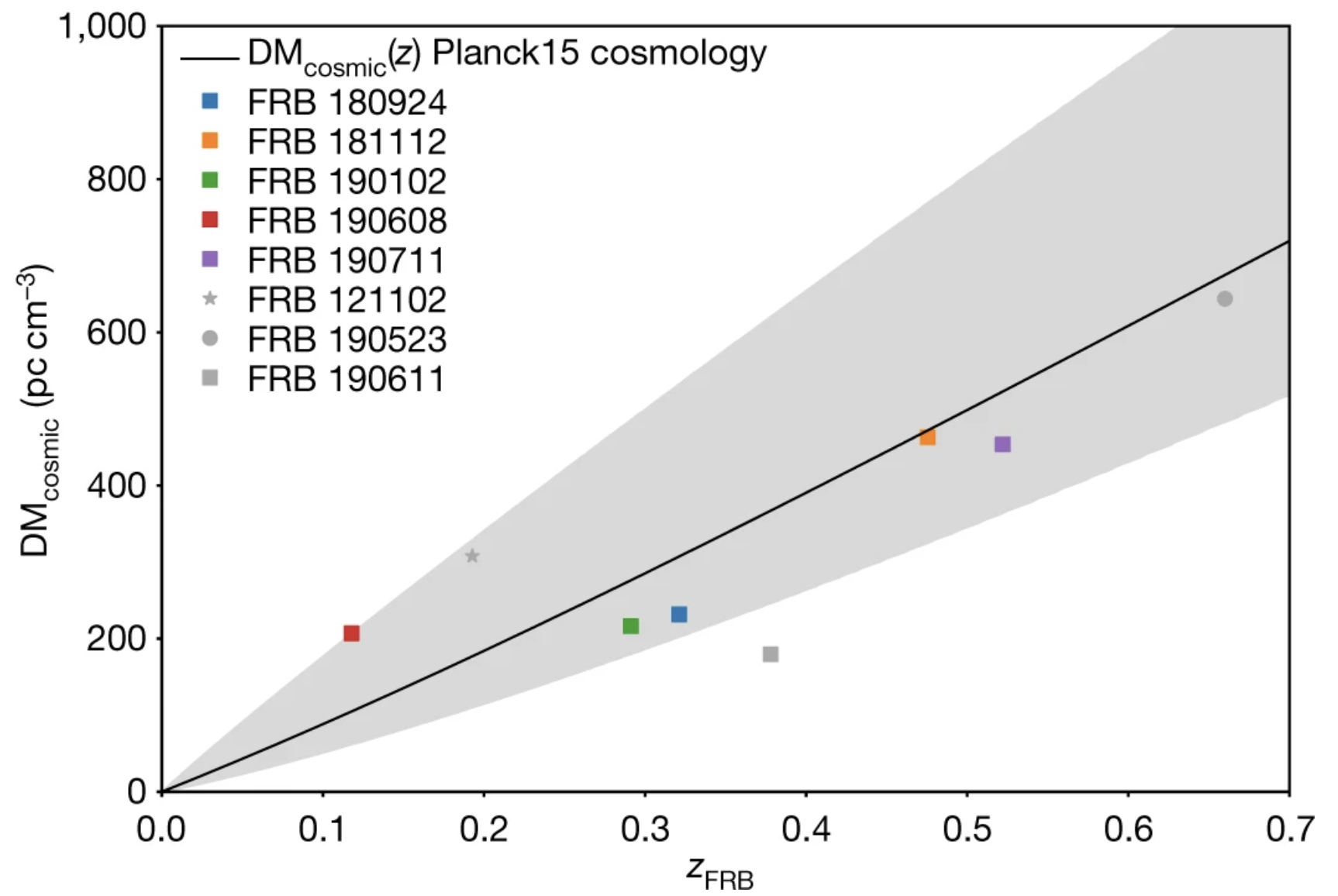}
\caption{The Macquart relation, as established for the first FRBs with robust host galaxy identifications and redshifts. The correlation between DM and redshift provides a compelling detection of baryons in the intergalactic medium. Credit: J-P Macquart.}\label{fig8}
\end{figure}

The scatter around the Macquart relation contains information about the turbulence of the interstellar gas around galaxy clusters and is now being studied in great detail~\citep{2023arXiv230507022B}. We anticipate a significant increase in the sample of FRBs with measured redshifts in the coming years. 

Among the many exciting results from individual sources that are now coming to light are the detections of thousands of bursts from selected FRB repeaters, in particular using FAST (Five Hundred Metre Aperture Spherical Telescope) in China~\citep{2021Natur.598..267L}. One very recent result by the FAST group shows that the bright FRB-like pulses from the Galactic magnetar appear at random rotational phases of the neutron star~\citep{2023SciA....9F6198Z}. The fainter radio pulsar-like pulses are seen only in a well defined rotational phase window. This potentially explains why, to date, no repeating FRB has shown a periodicity that is conclusively connected to a pulsar-like rotation.

One of the great debates that is currently ongoing in the FRB community is whether all FRBs repeat or not? As shown in the first CHIME catalogue, the morphological appearance of repeating versus one-off FRBs does appear to be very different: repeaters occupy only a fraction of the observable band, and produce broader pulses on average compared to the one of sources which tend to be narrower pulses that are broad band in nature~\citep{2021ApJ...923....1P}. Efforts to unify this dichotomy as some sort of geometrical effect (as has been proposed, for example, with active galactic nuclei), are challenging. All three of the authors of this paper have different opinions about the nature of repeating versus non-repeating FRBs!

\begin{figure}[h]
\centering
\includegraphics[width=0.8\textwidth]{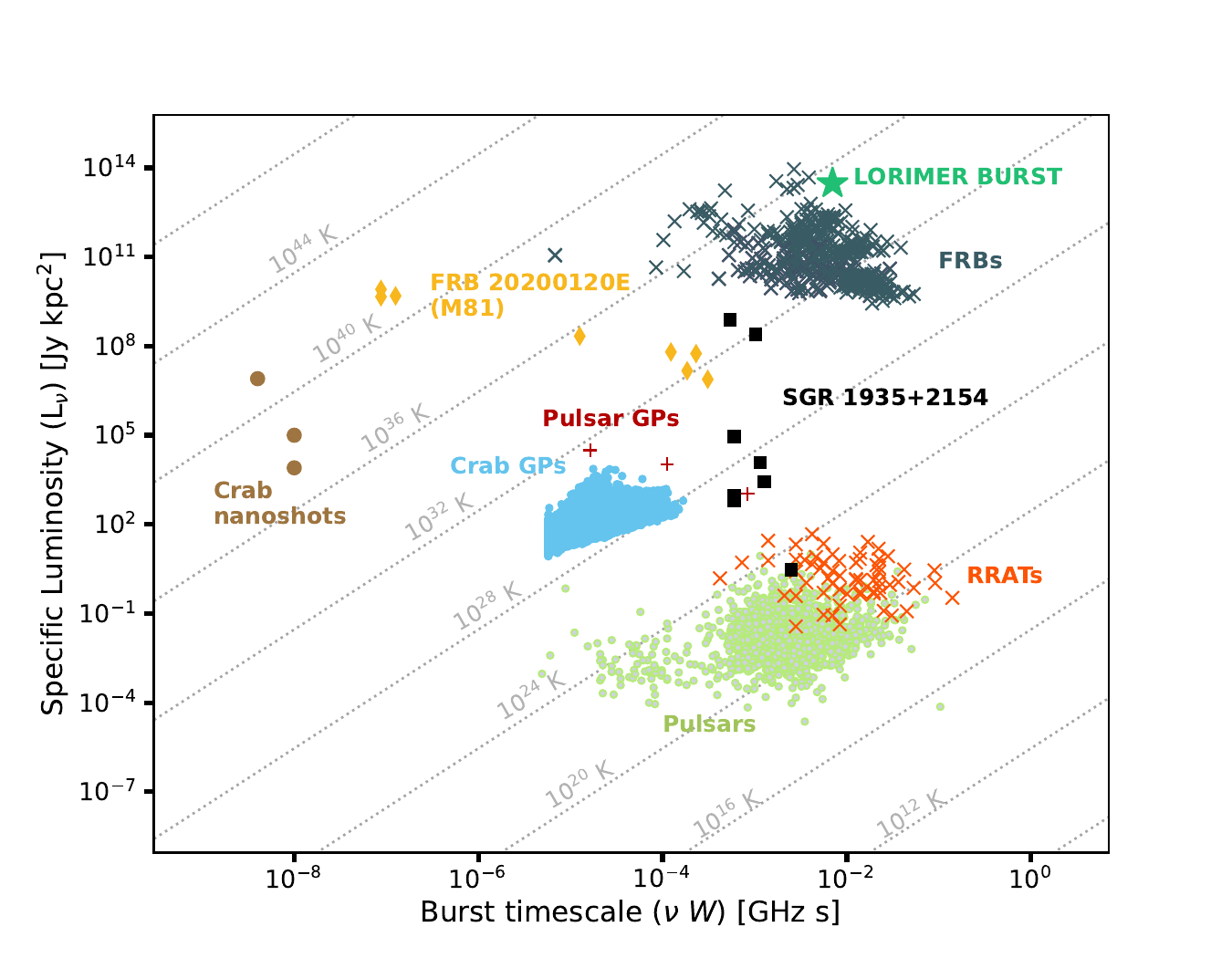}
\caption{Individual pulses from a variety of different sources displayed as luminosity versus pulse width times observing frequency. The Lorimer burst is no longer an anomalously bright object when compared to the other FRBs known. As we note in the text, there is still much parameter space to be explored. Further exciting discoveries in the transient radio sky are anticipated. Credit: Chris Flynn and Manisha Caleb.}\label{fig9}
\end{figure}

Regardless of all the mysteries that are still in play, one thing is very clear: as we look at the current state of the transient sky in terms of the “luminosity --- pulse duration” diagram shown in Fig.~\ref{fig9}, we see clear groupings of sources, but also lots of gaps in the parameter space.

These gaps indicate opportunities and challenges to the FRB community. We are not, for example, currently able to resolve pulses well enough to rule out a class of extremely short duration FRBs (dubbed ultrafast FRBs by some). Following predictions about a substantial population of ultra-long period Galactic magnetars \citep{2020MNRAS.496.3390B}, and
the fact that most searches to date are less sensitive to long-duration pulses, examples of long period (minutes or more) transients in the Galaxy are now being found \citep{2022Natur.601..526H,2023Natur.619..487H}. It is important to recognize these current shortcomings and opportunities as we plan future experiments.  Further discussion of this emerging population can be found in \citet{2023MNRAS.520.1872B}.

In closing, the future of FRBs is very exciting thanks to all of the developments that have taken place since our discovery in 2007. As mentioned above, many of the most recent results are challenging our understanding of the population which leads to a lot of open questions. We feel very fortunate to have played a role in sparking this whole community and continue to enjoy participating in it.

\bmhead{Acknowledgements}

We are grateful to a number of individuals for their roles in this work and organisations for supporting it over the last 17 years. Special thanks to Froney Crawford and Ash Narkevic for their contributions to the discovery. To the Shaw Foundation, and to those that nominate awardees and serve on our committees, thank you for highlighting the significance of the discovery of FRBs. Matthew has received significant support from Swinburne University of Technology and the Australian Research Council. Maura and Duncan are grateful for the sustained support from WVU, the National Science Foundation, the Research Corporation for Science Advancement and the West Virginia NASA Space Grant Consortium as well as the West Virginia Higher Education Policy Commission. We are grateful to our colleagues at the Australia Telescope National Facility, and the National Radio Astronomy Observatory for their support. Duncan would like to thank Enrico Massaro and Frank Drake for alerting him to some of the key references and developments concerning searches for extragalactic radio pulses prior to the discovery of FRBs, Paz Beniamini for highlighting the importance of long-period pulsars in the Galaxy and Ron Ekers for helpful comments.
{ We are very grateful to Fabio Santos and Elias Brinks at Springer for their patience as we put this manuscript together, and to the reviewer, Vicky Kaspi, for very useful feedback to an earlier draft of the manuscript.} Finally, huge thanks to all of our students and colleagues who have been a pleasure to work with and a constant source of inspiration.

\clearpage
\bibliography{refs}
\end{document}